\documentclass[prb,aps,showpacs]{revtex4}

\usepackage{graphicx}
\def\be{\begin{equation}}
\def\ee{\end{equation}}
\def\bea{\begin{eqnarray}}
\def\eea{\end{eqnarray}}
\def\ba{\begin{array}}
\def\ea{\end{array}}
\def\bdm{\begin{displaymath}}
\def\edm{\end{displaymath}}

\begin{document}

\title{Wavefunction topology of two-dimensional time-reversal symmetric superconductors}

\author{ S.-K. Yip}


\affiliation{Institute of Physics, Academia Sinica, Nankang, Taipei
115, Taiwan \\
e-mail: yip@phys.sinica.edu.tw}

\date{\today }

\begin{abstract}

We discuss the topology of the wavefunctions of two-dimensional
time-reversal symmetric superconductors.  We consider (a) the planar
state, (b) a system with broken up-down reflection symmetry, and (c)
a system with general spin-orbit interaction.  We show explicitly
how the relative sign of the order parameter on the two Fermi
surfaces affect this topology, and clarify the meaning of the $Z_2$
classification for these topological states. \\

Keywords: superconductivity, topological classifications

\end{abstract}

\pacs{74.20.Rp, 73.43.-f, 72.25.-b}


\maketitle

\section{Introduction}

Topological states of matter have received much recent attention.
\cite{Zhang-P} Best known examples are the two-dimensional integer
quantum Hall states.\cite{QH} These states can be characterized by
the Chern numbers associated with the electronic
wavefunctions.\cite{TKNN} At the boundary of the sample, the
non-trivial topological structure of these wavefunctions necessarily
implies the existence of chiral edge states.  These states can carry
charge (Hall) currents even though the bulk is gapped. Chern numbers
can be finite in systems only with broken time reversal symmetry,
which in this case is provided by the external magnetic field.
Recently, there is much interest in time-reversal symmetric
topological insulators.\cite{Kane05,Kane06,Kane07,Konig07,Hsieh08}
These insulators cannot be classified by the Chern numbers, yet they
are topologically distinct from ordinary insulators we learnt in
standard solid state textbooks.  The role of the $Z_2$ topological
number in these systems has been elucidated recently, especially by
Kane and his collaborators \cite{Kane05,Kane06,Kane07}. There are
also surface bound states at the edges of a topological insulator.
For a two-dimensional topological insulator, there exist a pair of
one-dimensional edge states, related by time-reversal symmetry, at
the boundary of the system. In the absence of spin-orbit
interaction, these edge states can be regarded as two
one-dimensionally dispersing states of opposite spins traveling with
opposite velocities.  These pair of states can carry spin current,
even though the bulk states are gapped.  These topologically
insulating states are often also referred to as spin-Hall
insulators.\cite{note3d}

There are directly analogies of topological states in
superconductors. Here, while the system possesses an extra mechanism
of transport via motion of Cooper pairs, the quasiparticle
wavefunctions can have topological properties imposed upon them by
the order parameter similar to those discussed above. Neither is one
 restricted to a periodic system; the
discussion can be done also in the continuum case.  A direct analogy
to the integer quantum Hall states are spinless
weak-coupling\cite{wc} superconductors where the order parameter are
$(k_x + i k_y)^n$, where $k_{x,y}$ are the components of the
wavevector $\vec k$ and $n$'s are (odd) integers
\cite{Volovik,Volovik97,Volovik99,Read00}. For the continuum, there
is no limit to the magnitude of $\vec k$. The Chern number is $n$.
For $n=1$, we have the familiar two-dimensional $p$ wave
superconductor which possesses a chiral gapless state at the edge of
the sample \cite{Volovik97,Read00}. The non-trivial topology of this
state is also reflected in the bound state spectra in a singly
quantized vortex\cite{Volovik99,Kopnin}. In this case, one finds a
Majorana mode at exactly zero energy. Majorana vortex bound states
and edge modes are intimately related to each other
\cite{Read00,Stone04}. The direct analogy of a time-reversal
symmetric topological insulator in two-dimension is the state often
referred to as the "planar state" in the $^3$He literature
\cite{Leggett}. Physically, this state is a "equal spin pairing
state",\cite{Leggett} where the up and down spins pair only among
themselves (but not with each other), with the down (up) spins
paired into a $(k_x + i k_y)$ ( $(k_x - i k_y)$) state. This state
is obviously time-reversal symmetric. Since the up and down spins
are decoupled, it is also clear that the boundary of such a
two-dimensional system would have one edge state for each spin
species, propagating in opposite directions \cite{Vorontsov08,Qi09}
(see also \cite{Junction09}).  A singly quantized vortex in the
planar state has two Majorana modes, one for each spin species. In
contrast, an s-wave superconductor is topologically trivial. It has
no topologically required edge states at the sample boundaries. No
Majorana mode exists in its vortex core, and the lowest energy bound
state has a finite energy \cite{deGennes}.

What is the fate of this superconducting state when the spins are no
longer decoupled?  Besides basic theoretical curiosity, this
question has now an added significance due to the study of
non-centrosymmetric superconductors.  These systems include three
dimensional systems such as $CePt_3Si$ and related compounds
\cite{Bauer07,Ce113}, or two-dimensional systems at the interface
between two different materials \cite{Reyren07}. Consider here in
particular a two-dimensional system without up-down reflection
symmetry, though still with rotation symmetry about the normal to
the plane.  In this system, the planar state is indistinguishable
from the s-wave pairing state by symmetry, so the two order
parameters can coexist \cite{Gorkov01}. However, an s-wave state is
topologically trivial, so there must be a non-trivial interplay
between the two order parameters when they are simultaneously
present.  In an earlier paper \cite{VBS}, we investigated the
presence or absence of zero energy modes at the vortex core for such
a state.  There we find that the zero energy modes exist only if the
p-wave order parameter is larger than the s-wave order parameter at
the Fermi surface.  In addition, we have included a Rashba energy
term in the single particle Hamiltonian, which spin-splits the Fermi
surface into two according to helicity. We concluded in this case
that the existence or absence of the zero energy states is
determined by the relative sign of the pairing order parameter on
these two Fermi surfaces.\cite{VBS}

The edge states at the boundary of a two dimensional
non-centrosymmetric superconductor have also been studied in
\cite{Tanaka09}.  There the authors have found the conditions on
existence or absence of edge states consistent with what we have
stated above. (see also \cite{Fujimoto09} for further discussions)

In this paper, motivated by the above, we try to understand the
results of ref \cite{VBS,Tanaka09} by directly examining the
topology of the quasiparticle wavefunctions in the bulk. A
fundamental concept in topological states is "obstruction", that is,
the inability to construct a wavefunction smooth in some parameter
space (here momentum space) satisfying certain basic criteria. For
the integer quantum Hall state, this has been illustrated explicitly
by Kohmoto \cite{Kohmoto85}.  There he showed that, for electrons in
a lattice and in the presence of a magnetic field (with rational
flux quanta per unit cell), one cannot have a Bloch wavefunction
that is smoothly defined throughout the entire (magnetic) Brillouin
zone. Dividing the momentum space into two regions, he then showed
that the transformation function relating these two regions (also
known as the "transition function") reflects the finite Chern number
of the state. Similar consideration has been utilized by Fu and Kane
(Appendix A of \cite{Kane06}):  an insulator is topologically
non-trivial if one cannot construct a pseudospin basis smooth within
the Brillouin zone (more precise statements below).  A
transformation function is then used to define the $Z_2$ topological
number.  We shall follow the same route here for our superconducting
states, with $\vec k$ however which can extend to infinity.   We
shall see
  directly how the relative sign of the order parameter on the Fermi
  surfaces
  determines whether it is possible or not for the wavefunction
  to be smooth in $\vec k$ space.  This thus in turn determines
  the $Z_2$ topological classification of the superconducting state.

  While preparing the present manuscript, a preprint \cite{Zhang09}
  on the topology of time-reversal symmetric weak-pairing superconductors in
  general dimensions appears,
  employing a method that is very different from ours.
  These authors started from the quasiparticle Hamiltonian rather than
  the wavefunctions.
  For three-dimensions, they developed their classification
  by employing a topological invariant introduced earlier
  \cite{Ludwig08}.  Their classification in two-dimensions is then
  obtained by performing a dimensional reduction from
  three-dimensions.
  For multi-Fermi surfaces, they showed that
  the product of the signs of the order parameter on
  Fermi surfaces enclosing time-reversal invariant points
  determines which $Z_2$ class the superconductor would belong.
  For a single pair of Fermi surfaces enclosing $\vec k = 0$, their
  result reduces to what we have stated above.  The
  arguments in Ref \cite{Zhang09}, though general,
   are in a rather abstract form.  For two-dimensions,
   they also rely on an artificial extension of
   the physical system to an extra dimension. We hope that
  some readers would find our explicit consideration here of obstruction
  and transformation function within the physical system
  more easily understandable and informative.

    Our paper is organized as follows.  For clearer explanations, we
    first start with the pure planar state, then introduce s-wave
    mixing and Rashba interaction.  We shall explicitly construct
    the $Z_2$ invariant.  Finally, we generalize to a general
    two-dimensional
    time-reversal symmetric state.  The Appendix contains some
    technical details.

    \section{Obstruction and the transformation matrix}

    \subsection{Planar phase}\label{sub:planar}

    We begin by considering a superconductor (superfluid) in the
    planar state.  The Hamiltonian is given by $H = H_K + H_{pair}$
    where the kinetic energy

    \be
    H_K = \sum_{\vec k, \alpha} \left( \frac{k^2}{2m} - \mu \right)
    a^{\dag}_{\vec k,\alpha} a_{\vec k, \alpha}
    \label{K}
    \ee

    \noindent and the pairing term $ H_{pair} \to
     H^{\rm planar}_{pair}$, with
    \be
    H^{\rm planar}_{\rm pair} = \frac{1}{2} \sum_{\vec k} \left(
    i \Delta_p(k) e^{ - i \phi_{\vec k}} a^{\dag}_{\vec
    k, \uparrow} a^{\dag}_{-\vec k, \uparrow} +
      i \Delta_p(k) e^{  i \phi_{\vec k}} a^{\dag}_{\vec
    k, \downarrow} a^{\dag}_{-\vec k, \downarrow}
    + h.c.   \right)
    \label{H-pair}
    \ee

    \noindent $k$, $\phi_{\vec k}$ are the
    magnitudes and azimuthal angles of $\vec k$, $\alpha =
    \uparrow, \downarrow$ the spins, $a_{\vec k,\alpha}$ is
    the corresponding annihilation operator, and $h.c.$ denotes the
    Hermitian conjugate. $m$ is the mass and $\mu$ is the chemical potential.
    $\Delta_p(k)$, which depends only on the
    magnitude of $\vec k$, is chosen real.  In eq (\ref{H-pair}),
    the up (down) spins have angular momentum $\mp 1$.
    To have a
    Hamiltonian smooth in $\vec k$ space, we shall require that $\Delta_p (k) \to 0$ as $k \to
    0$.   We shall
    adopt the following convention for the time reversal operator
    $\Theta$:  $ \Theta a^{\dag}_{\vec
    k, \uparrow} \Theta^{-1} = a^{\dag}_{- \vec
    k, \downarrow}$, and correspondingly
     $ \Theta a^{\dag}_{\vec k, \downarrow} \Theta^{-1} =
      - a^{\dag}_{- \vec k, \uparrow}$.  The Hamiltonian is time
      reversal invariant, for example
      $\Theta i e^{ - i \phi_{\vec k}} a^{\dag}_{\vec
    k, \uparrow} a^{\dag}_{-\vec k, \uparrow} \Theta^{-1}
     = i  e^{  i \phi_{\vec k}} a^{\dag}_{\vec
    k, \downarrow} a^{\dag}_{-\vec k, \downarrow}$.
    In principle we can also choose an $H^{\rm planar}_{\rm pair}$ which
    is time-reversal invariant under $\Theta$ only after an
    additional gauge transformation, but this would not add
    any new physics and we avoid such complications.  We could also
    have chosen a different $H^{\rm planar}_{\rm pair}$ such as
    $  \Delta_p(k) e^{ - i \phi_{\vec k}} a^{\dag}_{\vec
    k, \uparrow} a^{\dag}_{-\vec k, \uparrow} -
      \Delta_p(k) e^{  i \phi_{\vec k}} a^{\dag}_{\vec
    k, \downarrow} a^{\dag}_{-\vec k, \downarrow} $, but
    this would amount to replacing $\phi_{\vec k}$ in eq
    (\ref{H-pair}) by $\phi_{\vec k} - \frac{\pi}{2}$
    (that is, a relative spin-orbit rotation.  This state
    corresponds to the one used in \cite{Qi09}).  Thus
    eq (\ref{H-pair}) is the most general planar state.  The $i$
    factors in this equation are chosen for convenient comparison
    with the case with finite spin-orbit interactions
    (subsection B) below.

    This Hamiltonian can be easily diagonalized by Bogoliubov
    transformation, since the two spin species are decoupled.
    The energies are given by $\pm E(k)$ where
    $E(k) \equiv \sqrt{ \xi^2(k) + \Delta^2_p(k)}$, each doubly
    degenerate due to spin.  Here $\xi(k) = \frac{k^2}{2m} - \mu $.
    The unoccupied positive energy states \cite{unocc}
    $| \Psi^A_{\uparrow} (\vec k) > $,
    $| \Psi^A_{\downarrow} (\vec k) > $ have wavefunctions \cite{exp} given
    by, in the standard notation in spin and particle-hole space,

    \be
     \Psi^A_{\uparrow} (\vec k) = \mathfrak{N^A}
     \left( \ba{c} 1 \\ 0 \\
     - i \Delta_p(k)  e^{  i \phi_{\vec k}} / (E(k) + \xi(k)) \\
     0 \ea
     \right)
     \label{PsiA-u}
     \ee

     \be
     \Psi^A_{\downarrow} (\vec k) = \mathfrak{N^A}
     \left( \ba{c} 0 \\ 1 \\  0 \\
     - i \Delta_p(k)  e^{  - i \phi_{\vec k}} / (E(k) + \xi(k)) \\
      \ea
     \right)
     \label{PsiA-d}
     \ee
     Here $\mathfrak{N^A}$ is a normalization constant, which we
     choose to be real and thus
     $\mathfrak{N^A} \equiv \left( ( E(k) + \xi(k) )/ 2 E(k)
     \right)^{1/2}$.  For a given $\vec k$,
     $ |\Psi^A_{\uparrow} (\vec k)> $ and $|\Psi^A_{\downarrow} (\vec
     k)> $ form a complete set for the positive energy states of the
     Hamiltonian $H$.  In writing eqs (\ref{PsiA-u}) and
     (\ref{PsiA-d}) we have required that they satisfy the criteria
     for a good pseudospin pair as stated in Appendix of \cite{Kane06} (their eq (A1)),
     that is ,
     \be
     \Theta |\Psi_{1} (\vec k)> = |\Psi_{2} (-\vec k)>
     \label{ps1}
     \ee
     and correspondingly
     \be
     \Theta |\Psi_{2} (\vec k)> = - |\Psi_{1} (-\vec k)>
     \label{ps2}
     \ee
     when $1 \to \uparrow$ and $2 \to \downarrow$.\cite{T}

  We shall limit ourselves to the case where $\Delta_p(k)$ grows
  slower than $ k^2$ as $k \to \infty$, so that
  $ \Delta_p(k) /  ( E(k) + \xi(k) ) \to 0$ in this limit.
  Then
  $ |\Psi^A_{\uparrow} (\vec k)> $ and $|\Psi^A_{\downarrow} (\vec
     k)> $ are both well-defined as $\vec k \to \infty$
     \cite{infty}.  However, these wavefunctions are ill-defined
     at the origin $\vec k \to 0$ if the chemical potential $\mu$ is
     positive.  As $k \to 0$,
      $\Delta_p(k) / (E(k) + \xi(k)) = (E(k) -  \xi(k)) / \Delta_p (k)
      \to \infty$, since $\xi(k) \to - \mu < 0$ and $\Delta_p(k) \to
      0$.  Hence in this limit \cite{transpose}

      \bdm
      \Psi^A_{\uparrow} (\vec k) \to ( 0, 0, -i e^{ i \phi_{\vec
      k}}, 0)
      \edm
      \bdm
      \Psi^A_{\downarrow} (\vec k) \to ( 0, 0, 0, -i e^{ -i \phi_{\vec
      k}})
      \edm
      obviously ambiguous at $\vec k = 0$.

      We can however make the alternative choice,

       \be
     \Psi^B_{\uparrow} (\vec k) = \mathfrak{N^B}
     \left( \ba{c} i  \Delta_p(k)  e^{ - i \phi_{\vec k}} / (E(k) - \xi(k)) \\ 0 \\
      1 \\
     0 \ea
     \right)
     \label{PsiB-u}
     \ee

     \be
     \Psi^B_{\downarrow} (\vec k) = \mathfrak{N^B}
     \left( \ba{c} 0 \\
       i \Delta_p(k)  e^{   i \phi_{\vec k}} / (E(k) - \xi(k)) \\  0 \\
     1  \ea
     \right)
     \label{PsiB-d}
     \ee
      satisfying the requirements eqs (\ref{ps1}) and (\ref{ps2}),
       Here $\mathfrak{N^B} \equiv \left( ( E(k) - \xi(k) )/ 2 E(k)
     \right)^{1/2}$.  These wavefunctions are well-defined at the origin.
     As $k \to 0$, $\Delta_p(k) / (E(k) - \xi(k))
     \to 0$, so
     \bdm
      \Psi^B_{\uparrow} (\vec k) \to ( 0, 0, 1, 0)
      \edm
      \bdm
      \Psi^B_{\downarrow} (\vec k) \to ( 0, 0, 0 , 1)
      \edm
      However, they are ill-defined as $k \to
     \infty$.  In this limit,
      $\Delta_p(k) / (E(k) - \xi(k)) = (E(k) + \xi(k))/ \Delta_p(k)
      \to \infty$ so
      \bdm
      \Psi^B_{\uparrow} (\vec k) \to ( i e^{ - i \phi_{\vec k}}, 0, 0, 0)
      \edm
      \bdm
      \Psi^B_{\downarrow} (\vec k) \to ( 0, i e^{ i \phi_{\vec k}}, 0, 0 )
      \edm
      and have values which are $\phi_{\vec k}$
      dependent.\cite{infty}

      Thus $|\Psi^A_{\uparrow,\downarrow}(\vec k)>$ forms a good
      pseudospin basis
       for any $\vec k$ excluding the origin, while
      $|\Psi^B_{\uparrow,\downarrow}(\vec k)>$ is good
      for any $\vec k$ except at infinity.  To have a good
      pseudospin basis for all $\vec k$, one must divide the $\vec
      k$ space into two parts, say $k \ge K$ and $ 0 \le k \le K$ for
      some $K$, and use respectively $|\Psi^A_{\uparrow,\downarrow}(\vec k)>$ and
      $|\Psi^B_{\uparrow,\downarrow}(\vec k)>$ in these two regions.
      For our specific example, $K$ can assume any finite non-zero
      value.  At the boundary between these two regions (here $k =
      K$), we consider the unitary transformation matrix
      $\mathbf{t^{AB}}$  between
      these two basis (see Appendix A of
      \cite{Kane06}, though the details of our analysis is different
      from this reference; see also \cite{Roy06,Lee08}),
      with matrix elements defined by
      \be
      t^{AB}_{ij} (\phi) \equiv < \Psi^A_i(\phi) | \Psi^B_j(\phi)
      >_{k = K}
      \label{t-def}
      \ee
      which is now a function of the angle
      $\phi$ (with $K$ a parameter).
      The requirements eq (\ref{ps1}) (\ref{ps2}) impose
      a non-trivial constraint on this matrix.
      We have
      \be
       \left( | \Psi_1 (\phi+\pi) > , | \Psi_2 (\phi+\pi) >
       \right)_k
             =
       \left( |\Theta \Psi_1 (\phi) > , |\Theta \Psi_2 (\phi) >  \right)_k
               \left( \ba{cc}  0 & 1 \\
                              -1 & 0  \ea \right)
      \ee
       for both A and B and
       \be
       \mathbf{t^{AB}}(\phi + \pi) =
       \left( \ba{c} < \Psi^A_1 (\phi+\pi)| \\ <\Psi^A_2 (\phi+\pi)| \ea
       \right)_k
       \ba{c}
       \left( | \Psi^B_1 (\phi+\pi)> ,  |\Psi^B_1 (\phi+\pi)>
       \right)_k \\     \qquad     \ea
       \ee
       We therefore find
       \be
       \mathbf{t^{AB}} (\phi + \pi) =
       \left( \ba{cc} 0 & -1 \\
                              1 & 0  \ea \right)
             \mathbf{t^{AB}} (\phi)^{*}
       \left( \ba{cc} 0 & 1 \\
                              -1 & 0  \ea \right)
       \label{sym1}
       \ee
       Since $\mathbf{t^{AB}}$ is unitary, we can write
       $\mathbf{t^{AB}} = e^{ i \chi(\phi)} \mathbf{G} (\phi)$
       where $\mathbf{G}$ is an element in SU(2)
       (${\rm det} \mathbf{G} = 1$)\cite{ambi}  and $\mathbf{G(\phi)}$ is required
       to be smooth. Hence we have
       \be
          e^{ i \chi(\phi + \pi )} \mathbf{G} (\phi + \pi)
            =  e^{ - i \chi(\phi)} \mathbf{G} (\phi)
       \ee
       and so there are two possibilities, either
       \be
         e^{ i \chi(\phi + \pi )}
            =  e^{ - i \chi(\phi)} \qquad
            \mathbf{G} (\phi + \pi) =  \mathbf{G} (\phi)
         \label{trivial}
         \ee
         or
         \be
         e^{ i \chi(\phi + \pi )}
            =  - e^{ - i \chi(\phi)} \qquad
            \mathbf{G} (\phi + \pi) =  - \mathbf{G} (\phi)
         \label{non-trivial}
         \ee
         That is, if we consider the map from
         $\phi \in [ 0, \pi ]$ to $\mathbf{G}(\phi)$, the SU(2) part of $\mathbf{t^{AB}}$,
         it must be either periodic (eq (\ref{trivial})) or
         antiperiodic (eq (\ref{non-trivial})).  We note that, if
         this map is antiperiodic, there is no smooth deformation
         of the wavefunction which can turn it to be periodic, and
         so the periodic versus antiperiodic boundary condition can
         be used for a topological classification ($Z_2$).  Note
         also that once $\mathbf{G}(\phi)$ is given between $ 0$ and $\pi$,
         its value between $ \pi$ and $ 2 \pi$ would then be given.

         For our specific example, we can work out easily
         \be
         \mathbf{t^{AB}} (\phi) =
         \left( \ba{cc} i e^{ - i \phi} & 0 \\
                       0 & i e^{ i \phi} \ea \right)
                       \label{t-planar}
                       \ee
         and so
         \be
         \mathbf{G} (\phi) =
         \left( \ba{cc}  e^{ - i \phi} & 0 \\
                       0 & e^{ i \phi} \ea \right)
                       = e^{ - i \phi \mathbf{\tau_3}}
         \ee
         where $\mathbf{\tau_3} = \left( \ba{cc} 1 & 0 \\ 0 & -1 \ea
         \right)$,
         so $\mathbf{G}(\phi) $ is antiperiodic when $\phi \to \phi + \pi$.
         We reproduce the well-known result that our p-wave planar
         state is $Z_2$ non-trivial.

         Our result for $\mathbf{G}(\phi)$ can be easily understood,
         since the two spin species are decoupled.  The factors
         $ e^{ \mp i \phi}$ in its diagonal elements just reflect
         the $ \mp 1$ Chern numbers for the two spin species.
         The same calculation for an s-wave superconductor would
         show that $\mathbf{G}(\phi)$ is periodic
         (in fact a constant), hence topologically trivial.
         This example, though elementary, would be useful to
         understand the more complicated cases below.

 \subsection{Superconductor with Rashba interaction}
 \label{sub:Rashba}

 We now consider a superconducting  system without up-down
 reflection
 symmetry.  This system has now been studied much theoretically
 (e.g. \cite{Gorkov01,VBS,Tanaka09,Fujimoto09,Yip02,Curnoe04}).
  First let us consider the normal
 state.  In addition to the kinetic energy $H_K$ quadratic in $k$, we now
 add the Rashba (single-body) term
 \be
 H_R = \sum_{\vec k, \alpha, \beta}
  - \alpha_R(k)  a^{\dag}_{\vec k, \alpha} ( \hat z \times \hat k) \cdot \vec
 \sigma_{\alpha,\beta} a_{\vec k, \beta}
 \label{HR}
 \ee
 Here $\vec \mathbf{\sigma}$ are the Pauli spin matrices, and
 $\alpha_R(k) > 0$ is a measure of the strength of this Rashba "interaction".
 This term can be regarded as a $\vec k$ dependent magnetic field,
 pointing along $\hat z \times \hat k$, thus the system is still
 rotationally symmetric about its normal.
 We shall take $\alpha_R(k) \to 0 $ as $k \to 0$, so that
 $H_R$ is well-defined there.
 In the normal state, the eigen-energies can
 easily be seen to be $\xi_{\pm}(k) = \frac{k^2}{2m} \mp \alpha_R(k) -
 \mu$, according to whether the spin is parallel or anti-parallel to
 the effective field at $\vec k$, and we shall denote these two bands
 as $\pm$.  These two bands thus have different Fermi radii
 $k_{F \pm}$.
 The spin dependent part of the wavefunctions can be chosen as
 \bea
 | \vec k + > &=& \frac{1}{\sqrt{2}}
 \left( \ba{c} 1 \\ i e^{ i \phi_{\vec k}} \ea \right) \nonumber \\
  & &  \label{wf} \\
  | \vec k - > &=& \frac{1}{\sqrt{2}}
 \left( \ba{c}  i e^{ -i \phi_{\vec k}} \\ 1 \ea \right) \nonumber
 \eea

We note that
 $\Theta | \vec k + >  =  i e^{ -i \phi_{\vec k}} | - \vec k + >$
and $\Theta | \vec k - >  = - i e^{ i \phi_{\vec k}} | - \vec k - >
$ and thus $\Theta a^{\dag}_{\vec k +} \Theta^{-1} = i e^{ -i
\phi_{\vec k}} a^{\dag}_{-\vec k +}$ and  $\Theta a^{\dag}_{\vec k
-} \Theta^{-1} = - i e^{ -i \phi_{\vec k}} a^{\dag}_{-\vec k -}$.
We note that $|\vec k \pm>$ do not form a good pseudospin basis
since they do not satisfy eq (\ref{ps1}) and (\ref{ps2}).  We shall
refer $\pm$ as helicity indices.

For the superconducting state, in the limit where the pairing is
weak compared with the splitting between the two Fermi surfaces, it
is sufficient to consider only pairing terms on the same band
\cite{Yip02}.  Thus we can write

\be
    H_{\rm pair} = \frac{1}{2} \sum_{\vec k} \left(
    i \Delta_+ (k) e^{ - i \phi_{\vec k}} a^{\dag}_{\vec
    k +} a^{\dag}_{-\vec k +} -
      i \Delta_- (k) e^{  i \phi_{\vec k}} a^{\dag}_{\vec
    k - } a^{\dag}_{-\vec k - }
    + c.c. \right)
    \label{H-pair2}
    \ee

 $H_{\rm pair}$ is time reversal invariant if both $\Delta_{\pm}
 (k)$ are real, since we have, e.g.,
 $\Theta  i e^{ -i \phi_{\vec k}} a^{\dag}_{\vec k +} a^{\dag}_{-\vec k +} \Theta^{-1} =
 i e^{ -i \phi_{\vec k}}
a^{\dag}_{\vec k +}  a^{\dag}_{-\vec k +}$. The origin of the $e^{
\mp i \phi_{\vec k}}$ terms here should not be confused with those
in eq (\ref{H-pair}): they are the result of the choice eq
(\ref{wf}) and do not necessarily imply p-wave pairing
\cite{Gorkov01,Curnoe04}.  Using eq (\ref{wf}) we can rewrite
$H_{pair}$ in terms of ordinary spins and find \be
 H_{\rm pair} = \frac{1}{2} \sum_{\vec k, \alpha, \beta}  \left(
 a^{\dag}_{\vec k, \alpha} \Delta_{\alpha \beta} a^{\dag}_{-\vec k,
 \beta} + c.c. \right)
 \ee
 with
 \be
 \mathbf{\Delta(\vec k)} =
 \left( \ba{cc}    i \frac{\Delta_+(k) - \Delta_-(k)}{2} e^{ - i
 \phi_{\vec k}} &  \frac{\Delta_+(k) + \Delta_-(k)}{2} \\
 - \frac{\Delta_+(k) + \Delta_-(k)}{2}  &
   i \frac{\Delta_+(k) - \Delta_-(k)}{2} e^{  i
 \phi_{\vec k} }  \ea \right) \ee
 Thus it is a linear combination of s and p-wave planar pairing,
 with $\Delta_s (k) =  \frac{\Delta_+(k) + \Delta_-(k)}{2}$ and
 $\Delta_p (k) = \frac{\Delta_+(k) - \Delta_-(k)}{2}$.
 If $\Delta_+(k) = \Delta_-(k)$ for all $k$, we have pure s-wave pairing,
 and if $\Delta_+(k) = - \Delta_-(k)$, $H_{\rm pair}$ reduces
 to $H^{\rm planar}_{\rm pair}$.
 Generally $\Delta_{\pm} (k) = \Delta_s(k) \pm \Delta_p(k)$.  For
ease of discussions below, we shall always assume that $\Delta_+(k)$
is positive, where as $\Delta_-(k)$ may change sign as a function of
$k$.

  As $k \to \infty$, all pairing terms are negligible compared
  with the kinetic energy $k^2/2m$, and so the system is effectively
  normal.  To avoid the complications as mentioned below eq
  (\ref{wf}) we shall require further that $\alpha_R(k) \to 0$ there so
  that we can easily write down a good pseudospin pair
  \bea
  \Psi^{D1}_1
  &=& \left(  1,  0,  0, 0 \right) \nonumber \\
   & &   \label{D1}  \\
  \Psi^{D1}_2
  &=& \left(0, 1, 0, 0 \right) \nonumber
  \eea
  which is in fact the same as $|\mathbf{\Psi^A}>$ defined in
  the last subsection in this $k \to \infty$ limit.
 At $k \to 0$, since $\Delta_p$ vanishes, the pairing either vanishes completely,
 or is always s-wave like only.  Since $\alpha_R(k)$ there vanishes as well,
 we can have a pair of good pseudospin wavefunctions, which
 we write as
 \bea
 \Psi^{D2}_1 &=& \mathfrak{N^{D2}}\left(
 \ba{c} 0 \\ - \Delta_s(0)/ ( E(0) - \xi(0)) \\ 1 \\0 \ea \right)
 \nonumber \\
  & &   \label{D2}  \\
 \Psi^{D2}_2 &=& \mathfrak{N^{D2}}\left(
 \ba{c}  \Delta_s(0)/ ( E(0) - \xi(0)) \\ 0\\ 0 \\1 \ea \right)
 \nonumber
 \eea
  with $\mathfrak{N^{D2}} \equiv  (( E(0) - \xi(0))/ 2 E(0)
  )^{1/2}$.  Here $\xi(0) = \xi_{\pm}(0) =  - \mu$ is the kinetic energy as $k \to 0$,
  and $E(0) \equiv \sqrt{\xi^2(0) + \Delta_s^2(0)}$.

 For general $\vec k$, the Hamiltonian can be easily diagonalized in the $\pm$ helicity
basis.  The energies are $\pm E_{\pm}(k)$ with $E_{\pm}(k) = \sqrt{
\xi_{\pm}^2 (k) + \Delta_{\pm}^2 (k)}$.  The wavefunctions in
ordinary spin and particle hole space can then be found by the
transformation eq (\ref{wf}). We can choose
 \be
     \Psi^{C1}_{+} (\vec k) = \frac{\mathfrak{N_+^{C1}}}{\sqrt{2}}
     \left( \ba{c} 1 \\ i e^{  i \phi_{\vec k}} \\
     - i   e^{  i \phi_{\vec k}} \Delta_+(k)/ (E_+(k) + \xi_+(k)) \\
      \Delta_+(k) / (E_+(k) + \xi_+(k))\ea
     \right)
     \label{PsiC1+}
     \ee
     and
 \be
     \Psi^{C1}_{-} (\vec k) = \frac{\mathfrak{N_-^{C1}}}{\sqrt{2}}
     \left( \ba{c} i e^{  - i \phi_{\vec k}} \\  1 \\
     - \Delta_-(k) / (E_-(k) + \xi_-(k)) \\
      i   e^{ - i \phi_{\vec k}} \Delta_-(k)/ (E_-(k) + \xi_-(k))
      \ea
     \right)
     \label{PsiC1-}
     \ee
where $\mathfrak{N_{\pm}^{C1}}(k) = \left( ( E_{\pm}(k) +
\xi_{\pm}(k) )/ 2 E_{\pm}(k) \right)^{1/2}$ are normalization
constants. These functions are chosen so that they reduce to the
normal state wavefunctions eq (\ref{wf}) when $k \to \infty$.
 They are well-defined
for finite $k$ ($ 0 < k < \infty$) if $\Delta_-(k)$ does not change
sign anywhere for $k < k_{F-}$. (If $\Delta_-(k)$ changes sign at
$k^* < k_{F-}$, since $ \Delta_-(k)/ (E_-(k) + \xi_-(k)) = (E_-(k) -
\xi_-(k)) / \Delta_-(k)$ and $\xi_-(k^*)< 0$, $ (E_-(k) - \xi_-(k))
/ \Delta_-(k) \to \pm \infty$ according to whether $k$ approaches
$k^*$ from the $ {\rm sgn} \Delta_-(k) > 0$ or $< 0$ side.).
However, there is no harm if $\Delta_-(k)$ changes sign for $k >
k_{F-}$.

There is an alternative choice of wavefunctions
 \be
     \Psi^{C2}_{+} (\vec k) = \frac{\mathfrak{N_+^{C2}}}{\sqrt{2}}
     \left( \ba{c}  \Delta_+(k) / (E_+(k) - \xi_+(k)) \\
     i e^{  i \phi_{\vec k}}  \Delta_+(k) / (E_+(k) - \xi_+(k)) \\
     - i   e^{  i \phi_{\vec k}}\\ 1 \\
     \ea
     \right)
     \label{PsiC2+}
     \ee
     and
 \be
     \Psi^{C2}_{-} (\vec k) = \frac{\mathfrak{N_-^{C2}}}{\sqrt{2}}
     \left( \ba{c} i e^{  - i \phi_{\vec k}} \Delta_-(k)/ (E_-(k) - \xi_-(k)) \\
     \Delta_-(k) / (E_-(k) + \xi_-(k)) \\ -1\\
      i   e^{ - i \phi_{\vec k}}
      \ea
     \right)
     \label{PsiC2-}
     \ee
     where $\mathfrak{N_{\pm}^{C2}}(k) = \left( ( E_{\pm}(k) - \xi_{\pm}(k) )/ 2
    E_{\pm} (k) \right)^{1/2}$.  These wavefunctions are well-defined
    for $ 0 < k < \infty$ only if there are no sign changes of
    $\Delta_-(k)$ for $k > k_{F-}$.  (Actually $|\Psi_+^{C1}>$ and
    $|\Psi_+^{C2}>$ are always identical, since we assumed
    $\Delta_+(k) > 0$.  We just write explicitly
    $|\Psi_+^{C2}>$ again for symmetric purposes.)

    To study the topological classification as in subsection \ref{sub:planar},
    we need to construct good pseudospin bases satisfying eq
    (\ref{ps1}) and (\ref{ps2}) for different parts of $\vec k$ space,
     and then try to match them at the boundaries.
    However,  $|\Psi_{\pm}^{C1,C2}>$ do
    not form good pseudospin bases.  The problem is that,
    as in the normal state, the time reversed partner of the $+$
    band
    is the $+$ band itself.  Therefore to have a good pseudospin
    basis one must have $|\Psi_1 (\vec k)> $ proportional
    to $| \Psi_{+} (\vec k)>$ over part of the $\vec k$ space
    whereas it is proportional to $| \Psi_{-} (\vec k) >$ in the other,
    introducing discontinuities in the wavefunctions.  Rather than
    trying to deal with this rather messy situation, we note that
    the topological classification must be independent of the
    specific construction of the pseudospin basis.  If one has
    a smoothly defined wavefunction, such as $|\Psi_{\pm}^{C1,C2}>$
    above under suitable circumstances, then we should be able to
    decide the $Z_2$ invariant from these wavefunctions alone.
    We propose that this can done by studying the $SU(2)$ part of
    the matrix
    \be
    \mathbf{T}(\phi) \equiv < \mathbf{\Psi^{D1}} (\phi) | \mathbf{\Psi^C} (\phi) >_{k \to \infty}
            < \mathbf{\Psi^C} (\phi) | \mathbf{\Psi^{D2}} (\phi) >_{k \to 0}
            \label{T1}
    \ee
    with the matrix elements of each matrix defined as in eq
    (\ref{t-def}).
    Here $1 \to +$, $2 \to -$ for $|\Psi_{\pm}^C>$, and $C \to C1$
    or $C2$
     should be chosen so that it is smooth for $ 0 < k < \infty$.
     Intuitively, we are trying to consider the transformation
     between the good pseudospin bases $|\mathbf{\Psi^{D1}}>$
     at $k \to \infty$ and $|\mathbf{\Psi^{D2}}>$ at $k \to 0$ through
     intermediate states $|\mathbf{\Psi_{\pm}^C}>$ which are smoothly defined for
     $ 0 < k < \infty$.  Eq (\ref{T1}) is a special case of a more general formula
     (\ref{TG}) which
     we shall explain in the Appendix.

     We now evaluate eq (\ref{T1}).  Let us first
     consider the case where $\Delta_-(k)$ never changes sign for
     $k < k_{F-}$, so $|\mathbf{\Psi^{C1}_{\pm}}>$ can be used in eq
     (\ref{T1}).  We obtain
     \be
     < \mathbf{\Psi^{D1}} | \mathbf{\Psi^{C1}}> _{k \to \infty}
      = \frac{1}{\sqrt{2}}
      \left( \ba{cc} 1 & i e ^{ - i \phi} \\
              i e ^{ i \phi} & 1 \ea \right)
      \label{D1C1}
      \ee
      independent of the sign of $\Delta_{\pm} (k)$.
      For $k \to 0$, we get
      \be
     < \mathbf{\Psi^{D2}} | \mathbf{\Psi^{C1}}> _{k \to 0}
      = \frac{1}{\sqrt{2}}
      \left( \ba{cc} -i e ^{  i \phi}  & - {\rm sgn} \Delta_-(k_{F-}) \\
              1 & i e ^{ -i \phi} {\rm sgn} \Delta_-(k_{F-}) \ea \right)
      \ee
      Here we have used the fact that ${\rm sgn} \Delta_-(k \to 0)
      = {\rm sgn} \Delta_-(k_{F-})$ since $\Delta_-(k)$ never
      changes sign for $k < k_{F-}$.  Thus, for ${\rm sgn}
      \Delta_-(k_{F-}) > 0$, we get
      \bea
      \mathbf{T}(\phi) &=& \frac{1}{2}
       \left( \ba{cc} 1 & i e ^{ - i \phi} \\
              i e ^{ i \phi} & 1 \ea \right)
         \left( \ba{cc} i e ^{ - i \phi}  & 1 \\
              - 1 & - i e ^{ i \phi}  \ea \right)
              \nonumber \\
              &=&
              \left( \ba{cc} 0 & 1 \\ -1 & 0 \ea \right)
        \label{a1}
        \eea
        independent of $\phi$.  This state is $Z_2$ trivial.
        For ${\rm sgn} \Delta_-(k_{F-}) < 0$, we get instead
      \bea
      \mathbf{T}(\phi) &=& \frac{1}{2}
       \left( \ba{cc} 1 & i e ^{ - i \phi} \\
              i e ^{ i \phi} & 1 \ea \right)
         \left( \ba{cc} i e ^{ - i \phi}  & 1 \\
               1 &  i e ^{ i \phi}  \ea \right)
              \nonumber \\
              &=&
              \left( \ba{cc} i e ^{ - i \phi}  & 0 \\  0  & i e ^{i \phi} \ea
              \right)  \qquad ,
        \label{a2}
        \eea
        the same as eq (\ref{t-planar}), the result for the planar phase in the last
        subsection.   This state is therefore $Z_2$ non-trivial.  We
        note that the above calculation applies even when
        $\Delta_-(k)$ changes sign for some $k > k_{F-}$.

        For the case where $\Delta_-(k)$ changes sign for $k <
        k_{F-}$ but not for $k > k_{F-}$, we must employ instead
        $|\mathbf{\Psi^{C2}_{\pm}}>$.  We find
          \be
        < \mathbf{\Psi^{D1}} | \mathbf{\Psi^{C2}}> _{k \to \infty}
      = \frac{1}{\sqrt{2}}
      \left( \ba{cc}  1 & i e ^{ -i \phi} {\rm sgn} \Delta_-(k_{F-}) \\
              i e ^{ i \phi} & {\rm sgn} \Delta_-(k_{F-}) \ea \right)
      \ee
      where we have used $ {\rm sgn} \Delta_-(k_{F-}) = {\rm sgn} \Delta_-(k \to
      \infty)$.
      On the other hand,
      \be
     < \mathbf{\Psi^{D2}} | \mathbf{\Psi^{C2}}> _{k \to 0}
      = \frac{1}{\sqrt{2}}
      \left( \ba{cc}- i e ^{  i \phi} & - 1 \\
             1 &  i e ^{ -i \phi} \ea \right)
      \label{D2C2}
      \ee
      independent of the sign of $\Delta_-(k_{F-})$.
      Hence, if  ${\rm sgn} \Delta_-(k_{F-}) >  0$, we get
     \bea
      \mathbf{T}(\phi) &=& \frac{1}{2}
       \left( \ba{cc} 1 & i e ^{ - i \phi} \\
              i e ^{ i \phi} & 1 \ea \right)
         \left( \ba{cc} i e ^{ - i \phi}  & 1 \\
              - 1 & - i e ^{ i \phi}  \ea \right)
              \nonumber \\
              &=&
              \left( \ba{cc} 0 & 1 \\ -1 & 0 \ea \right)
        \label{b1}
        \eea
        as in eq (\ref{a1}), independent of $\phi$.  This state is $Z_2$ trivial.
        If  ${\rm sgn} \Delta_-(k_{F-}) <   0$ instead,
        we get
       \bea
      \mathbf{T}(\phi) &=& \frac{1}{2}
       \left( \ba{cc} 1 & - i e ^{ - i \phi} \\
              i e ^{ i \phi} & - 1 \ea \right)
         \left( \ba{cc} i e ^{ - i \phi}  & 1 \\
               -1 &  - i e ^{ i \phi}  \ea \right)
              \nonumber \\
              &=&
              \left( \ba{cc} i e ^{ - i \phi}  & 0 \\  0  & i e ^{i \phi} \ea
              \right)  \qquad ,
        \label{b2}
        \eea
        as in eq (\ref{b2}).  This state is again $Z_2$ non-trivial.

        We thus conclude that the system is $Z_2$ trivial or non-trivial
        according to ${\rm sgn} \Delta_-(k_{F-}) > (<)    0$,
        correspondingly whether $\Delta_s(k) > (<) \Delta_p(k)$ at
        $k_{F-}$, a criterion which we had obtained before by
        considering the vortex bound states \cite{VBS}.
        This result can also be understood from continuity.  When
        changing $\Delta_-(k_{F-})$, from say, positive to negative,
        one must go through the point where
        $\Delta_-(k_{F-}) = 0$, which indicates
        a gapless phase since then
        $E_-(k_{F-}) = \sqrt{\xi_-^2 (k_{F-}) + \Delta_-^2 (k_{F-})}$
        would vanish on the Fermi surface $\xi_-(k_{F-}) = 0$
        \cite{VBS}.
        Hence if $\Delta_-(k_{F+})$ and $\Delta_-(k_{F-})$ are of
        the same sign, then one can smoothly change
        $\Delta_{\pm}(k)$ to obtain the s-wave state $\Delta_+(k)
        =\Delta_-(k)$ without going through any gapless phase.
        Similarly,  if $\Delta_-(k_{F+})$ and $\Delta_-(k_{F-})$ are of
        the opposite sign, then the state is connected to planar state $\Delta_+(k)
        = - \Delta_-(k)$.

        Fu and Kane \cite{Kane05,Kane06} also proposed that studying the
        matrix elements (more precisely, the {\it Pfaffian})
         of $\Theta$ between wavefunctions at the same
        $\vec k$ point can distinguish whether the system is $Z_2$
        trivial or non-trivial.  For the system with only two
        (un-)occupied states $|\Psi_{1,2}(\vec k)>$, this just
        amounts to studying the function
        $P(\vec k) = < \Psi_{2}(\vec k)|\Theta |\Psi_{1}(\vec k)> $.
        The non-trivial topology of the system is determined by the
        presence of odd numbers of pairs of point zeros of $P(\vec k)$.  Since
        this overlap is evaluated at a given $\vec k$ (not relating $\pm \vec k$)
        and a
        transformation of basis would not affect the presence or
        absence of zeros of $P(\vec k)$, $|\Psi_{1,2}(\vec k)>$
        here need not be well-defined pseudospin states in the
        sense of eq (\ref{ps1}) and (\ref{ps2}).  From our previous
        calculations, we can also see in our example how $P(\vec k)$
        can distinguish $Z_2$ trivial versus non-trivial cases,
        provided one can use a smooth wavefunction in $\vec k$ for
        $|\Psi_{1,2}(\vec k)>$ (Here $|\Psi_{\pm}^{C1,C2}(\vec
        k)>$),
        except here that zeros of $P(k)$ must now appear as circles.
        Suppose that $\Delta_-(k)$ never changes sign for $k <
        k_{F-}$, then we can employ $|\Psi_{\pm}^{C1}(\vec k)>$.
        We find
        \bea
        P(k) &=& <\Psi_{-}^{C1}(\vec k) |\Theta |\Psi_{+}^{C1}(\vec k)>
         \nonumber \\
          &=& \mathfrak{N^{C1}}_+ \mathfrak{N^{C1}}_- \
          [ 1 + \frac{\Delta_+(k)}{ E_+(k) + \xi_+(k)}
                \frac{\Delta_-(k)}{ E_-(k) + \xi_-(k)} ]
                \eea

       As $k \to \infty$, $P(k) = 1$, but as $k \to 0$,
       $P(k) \to {\rm sgn} \Delta_+(0)  {\rm sgn} \Delta_-(0)
         = {\rm sgn} \Delta_-(0) = {\rm sgn} \Delta_-(k_{F-})$ \cite{even}.
         If $\Delta_-(k_{F-}) > 0$, then $P(0) = 1$.  However,
         if $\Delta_-(k_{F-}) < 0$, then $P(0) = -1$, implying
         $P(k)$ must vanish at some intermediate $k$ since
         $ | \Psi_{\pm}^{C1}(\vec k)>$ are smooth.  In the special
         case $k_{F+} = k_{F-}= k_F$, this zero occurs exactly
         at $k_F$, since there $\xi_+(k) = \xi_-(k) = 0$ and so
         $P (k_F) = [ 1 +  {\rm sgn} \Delta_+(k_{F}) {\rm sgn}
         \Delta_-(k_{F})] / 2$.  If $k_{F+}$ and $k_{F-}$ are not
         equal, it can shown that $P(k)$ vanishes somewhere between
         these two Fermi surfaces.

       Our calculations above do not apply if $\Delta_-(k)$ changes
       sign for both $k$ $>$ and $<$ $k_{F-}$.  If this happens,
       a further generalization of eq (\ref{T1}) is necessary.
       We can now write

       \be
       \mathbf{T}(\phi)
 = < \mathbf{\Psi^{D1}}(\phi) | \mathbf{\Psi^{C1}}(\phi) >_{\infty}
 < \mathbf{\Psi^{C1}}(\phi) | \mathbf{\Psi^{C2}}(\phi) >_{k_{F-}}
  < \mathbf{\Psi^{C2}}(\phi) | \mathbf{\Psi^{D2}}(\phi) >_{0}
       \label{TR2}
       \ee

       This equation is a special case of eq (\ref{TG2}) in the
       Appendix. Using eq (\ref{D1C1}), (\ref{D2C2}) and

       \be
       < \mathbf{\Psi^{C1}}(\phi) | \mathbf{\Psi^{C2}}(\phi) >_{k_{F-}}
       = \left( \ba{cc} 1 & 0 \\ 0 & {\rm sgn} \Delta_(k_{F-}) \ea
       \right)
       \ee
       we can easily check that the SU(2) part of $\mathbf{T}$ is
       periodic or antiperiodic in $\phi \to \phi + \pi$ depending
       on whether $ {\rm sgn} \Delta_-(k_{F-}) > (<) 0$.

  \subsection{General spin-orbit interaction}\label{sub:so}

  We generalize the above to general spin-orbit interaction
  \be
  H_{so} =  \sum_{\vec k, \alpha, \beta}
  -   a^{\dag}_{\vec k, \alpha} \vec h_{\vec k}  \cdot \vec
 \sigma_{\alpha,\beta} a_{\vec k, \beta}
 \label{H-so}
 \ee
 We parameterize the direction $\hat h (\vec k)$ by the spherical
 angles $(\beta_{\vec k}, \alpha_{\vec k})$, and thus we have
 $ \vec h_{\vec k} = | \vec h_{\vec k} |
 ( {\rm sin} \beta_{\vec k} {\rm cos} \alpha_{\vec k},
    {\rm sin} \beta_{\vec k} {\rm sin} \alpha_{\vec k},
    {\rm cos} \beta_{\vec k} )$ in Cartesian coordinates.
    Time reversal invariance requires
    \bea
    |\vec h (\phi+\pi, k)| &=& |\vec h (\phi, k)| \nonumber \\
    \beta(\phi+\pi, k) &=& \pi - \beta (\phi, k) \label{tr} \\
    \alpha(\phi+\pi, k) &=& \alpha (\phi, k) + \pi
    \qquad ({\rm mod}\ 2 \pi)  \nonumber
    \eea
    but we shall otherwise allow  general $\vec k$ dependence for
    these quantities (cylindrical symmetry would not be enforced).
    $|\vec h_{\vec k}|$ will be required to vanish at $\vec k = 0$,
    and we shall assume that it is finite for $ 0 < k < \infty$
    so that the bands are non-degenerate.  To have well-defined
    $\alpha_{\vec k}$, $\beta_{\vec k}$ cannot be equal to $0$ or
    $\pi$.  We shall assume that $\hat h_{\vec k}$ does not
    cover the entire sphere, so with suitable rotations of
    the spin quantization axes, this condition can be fulfilled.
    Otherwise some extra care has to be exercised when using
    the formulas below, but we shall not go into these complications
    here.  If $|\vec h_{\vec k}|$ is independent of directions,
    and further $\beta_{\vec k} = \pi/2$, $\alpha_{\vec k} = \phi_{\vec k} +
    \pi/2$, then (\ref{H-so}) reduces to the Rashba interaction
    (\ref{HR}).

    For the normal state, the eigenfunctions, with energies
    $\xi_{\pm}(\vec k) = \frac{k^2}{2m} \mp |\vec h_{\vec k}| - \mu$,
     can be chosen to be
    \bea
    | \vec k + > &=&  \left(
    \ba{c}  {\rm cos} (\beta_{\vec k}/2) \nonumber \\
      {\sin} (\beta_{\vec k}/2) e^{ i \alpha_{\vec k}} \ea \right)
      \\
       & &  \label{wf2} \\
   | \vec k - > &=&   \left(
    \ba{c} -{\rm sin} (\beta_{\vec k}/2) e^{ - i \alpha_{\vec k}} \\
      {\rm cos} (\beta_{\vec k}/2)  \ea \right) \nonumber
      \eea
    generalizing eq (\ref{wf}).
    We now have the time-reversal relations
    $\Theta | \vec k + >  =  - e^{ -i \alpha_{\vec k}} | - \vec k + >$
and $\Theta | \vec k - >  = - e^{ i \alpha_{\vec k}} | - \vec k -
>$.  The angular dependent Fermi surfaces will be denoted
 as $k_{F \pm } (\phi)$.

In weak pairing limit so that the pairing is only between degenerate
bands, we have

\be
   H_{\rm pair} = \frac{1}{2} \sum_{\vec k} \left(
    - \Delta_+ (\vec k) e^{ - i \alpha_{\vec k}} a^{\dag}_{\vec
    k +} a^{\dag}_{-\vec k +} -
       \Delta_- (\vec k) e^{  i \alpha_{\vec k}} a^{\dag}_{\vec
    k - } a^{\dag}_{-\vec k - }
    + c.c. \right)
    \label{H-pair3}
    \ee
    generalizing eq (\ref{H-pair2}).  Here $\Delta_{\pm}(\vec k)$
    are real.  As before, the $e^{ \mp i \alpha_{\vec k}}$ phase
    factors are due to the wavefunctions $|\vec k \pm>$.  $H_{pair}$
    obeys time-reversal invariance.  It is a mixture of singlet and
    triplet pairs, but the later is not necessarily a simple planar
    state.  We shall take $\Delta_+(\vec k) > 0$ always, but
    $\Delta_-(\vec k)$ can take either sign.  If the system is to
    remain gapped, then $\Delta_- (k_{F-} (\phi))$ must be either
    be all positive or all negative for all angles $\phi$,
    a situation which we shall assume.
    Note that as $\vec k
    \to 0$, all components of the p-wave pairing must go to zero,
    with only the s-wave component $\Delta_s$ surviving.
    There, $\Delta_+(0) = \Delta_-(0) = \Delta_s (0)$.

    In the limit $\vec k \to \infty$ ($0$), we again have the good
    pseudospin basis $|\mathbf{\Psi^{D1}}>$ ($|\mathbf{\Psi^{D2}}>$)
    as in eq (\ref{D1}) (eq (\ref{D2})).  For $ 0 < k < \infty$ we can
    solve the Hamiltonian again first in the $\pm$ helicity basis,
    then perform the transformation using (\ref{wf2}).  These
    wavefunctions can be written as
    \be
     \Psi^{C1}_{+} (\vec k) = \mathfrak{N_+^{C1}}
     \left( \ba{c} {\rm cos} (\beta_{\vec k}/2) \\
      {\rm sin} (\beta_{\vec k}/2) e^{  i \alpha_{\vec k}} \\
     - {\rm sin} (\beta_{\vec k}/2)   e^{  i \alpha_{\vec k}}
     \Delta_+(\vec k)/ (E_+(\vec k) + \xi_+(\vec k)) \\
      {\rm cos} (\beta_{\vec k}/2)
      \Delta_+(\vec k) / (E_+(\vec k) + \xi_+(\vec k))\ea
     \right)
     \label{PsiCC1+}
     \ee
     and
 \be
     \Psi^{C1}_{-} (\vec k) = \mathfrak{N_-^{C1}}
     \left( \ba{c} - {\rm sin} (\beta_{\vec k}/2) e^{  - i \alpha_{\vec k}} \\
       {\rm cos} (\beta_{\vec k}/2) \\
     - {\rm cos} (\beta_{\vec k}/2)
      \Delta_-(\vec k) / (E_-(\vec k) + \xi_-(\vec k)) \\
      - {\rm sin} (\beta_{\vec k}/2)   e^{ - i \alpha_{\vec k}}
      \Delta_-(\vec k)/ (E_-(\vec k) + \xi_-(\vec k))
      \ea
     \right)
     \label{PsiCC1-}
     \ee
     These wavefunctions are smooth in $\vec k$ if $\Delta_-(\vec k)$ do not
     change sign for $\vec k$ inside the $k_{F-}$ Fermi surface ($\mu > 0$).

     The alternative choice of wavefunctions
     \be
     \Psi^{C2}_{+} (\vec k) = \mathfrak{N_+^{C2}}
     \left( \ba{c}  {\rm cos} (\beta_{\vec k}/2)
      \Delta_+(\vec k) / (E_+(\vec k) - \xi_+(\vec k)) \\
     {\rm sin} (\beta_{\vec k}/2) e^{  i \alpha_{\vec k}}
       \Delta_+(\vec k) / (E_+(\vec k) - \xi_+(\vec k)) \\
     - {\rm sin} (\beta_{\vec k}/2)   e^{  i \alpha_{\vec k}}\\
      {\rm cos} (\beta_{\vec k}/2) \\
     \ea
     \right)
     \label{PsiCC2+}
     \ee
     and
 \be
     \Psi^{C2}_{-} (\vec k) = \mathfrak{N_-^{C2}}
     \left( \ba{c}
     - {\rm sin} (\beta_{\vec k}/2)  e^{  - i \alpha_{\vec k}}
     \Delta_-(\vec k)/ (E_-(\vec k) - \xi_-(\vec k)) \\
     {\rm cos} (\beta_{\vec k}/2)
     \Delta_-(\vec k) / (E_-(\vec k) - \xi_-(\vec k)) \\
     -   {\rm cos} (\beta_{\vec k}/2) \\
      - {\rm sin} (\beta_{\vec k}/2)   e^{ - i \alpha_{\vec k}}
      \ea
     \right)
     \label{PsiCC2-}
     \ee
     are smooth if $\Delta_-(\vec k)$ does not change signs for
      $\vec k$ outside the $k_{F-}$ Fermi surface.

  To study the topology, we consider the matrix
   \be
    \mathbf{T}(\phi) \equiv < \mathbf{\Psi^{D1}} (\phi) | \mathbf{\Psi^C} (\phi) >_{k \to \infty}
    e^{ - i \Delta \alpha (\phi) \mathbf{\tau_3} / 2}
            < \mathbf{\Psi^C} (\phi) | \mathbf{\Psi^{D2}} (\phi) >_{k \to 0}
            \label{T2}
    \ee
   again as the transformation between $ |  \mathbf{\Psi^{D1}}>$ and
   $  |\mathbf{\Psi^{D2}}>$ via  smooth wavefunctions
   ($\mathbf{C} \to \mathbf{C1}$ or $\mathbf{C2}$) defined in the
   intermediate $\vec k$ region.  The factor
   $ e^{ - i \Delta \alpha (\phi)\mathbf{\tau_3}/2 }$ is to take care
   of the variation of the wavefunction
   $ | \mathbf{\Psi^C} (\phi) > $ from $k \to 0$ to $k \to \infty$,
   here $\Delta \alpha (\phi) \equiv \alpha(\phi,\infty) -
   \alpha(\phi,0)$.  For the justification of this formula, see the
   Appendix.

   Going through algebra similar to the last subsection,
   we find, if $\Delta_-(k_{F-}(\phi)) > 0$,
  \be
  \mathbf{T}(\phi) = \left(
  \ba{cc}   {\rm sin} (\frac{\beta_\infty - \beta_0}{2})
  e ^ { - i (\alpha_\infty + \alpha_0)/2} &
   {\rm cos} (\frac{\beta_\infty - \beta_0}{2})
  e ^ { - i \Delta \alpha/2} \\
  - {\rm cos} (\frac{\beta_\infty - \beta_0}{2})
  e ^ { i \Delta \alpha/ 2} &
    {\rm sin} (\frac{\beta_\infty - \beta_0}{2})
  e ^ { i (\alpha_\infty + \alpha_0)/2}
  \ea  \right)
  \label{t31}
  \ee
  Here $\beta_\infty \equiv \beta(\phi,\infty)$ etc,
  and we have suppressed the $\phi$ labels in eq (\ref{t31})
  to shorten notations.
  For the Rashba spin-orbit interaction, eq (\ref{t31})
  reduces to
  $ \left( \ba{cc} 0 & 1 \\ -1 & 0 \ea \right) $ we have met before.
  For here, we note that ${\rm det} \mathbf{T} = 1$, so that the
  $SU(2)$ part of $\mathbf{T}$ can be simply taken as $\mathbf{T}$
  itself.
  Generally, with eq (\ref{tr}), we see that $\mathbf{T}(\phi)$ is
  periodic when $\phi \to \phi + \pi$, as
  $\Delta \alpha (\phi + \pi) = \Delta \alpha(\phi)$ (see Appendix),
  and both ${\rm sin} (\frac{\beta_\infty - \beta_0}{2})$
  and $e ^ { \pm i (\alpha_\infty + \alpha_0)/2}$ change sign.
  This state is therefore topologically trivial.

  If $\Delta_-(k_{F-} (\phi)) <  0$, we find instead

  \be
  \mathbf{T}(\phi) = \left(
  \ba{cc}   - {\rm sin} (\frac{\beta_\infty + \beta_0}{2})
  e ^ { - i (\alpha_\infty + \alpha_0)/ 2} &
   {\rm cos} (\frac{\beta_\infty + \beta_0}{2})
  e ^ { - i \Delta \alpha/2} \\
   {\rm cos} (\frac{\beta_\infty + \beta_0}{2})
  e ^ { i \Delta \alpha/ 2} &
    {\rm sin} (\frac{\beta_\infty + \beta_0}{2})
  e ^ { i (\alpha_\infty + \alpha_0)/2}
  \ea  \right)
  \label{t32}
  \ee
  If $\beta_{\vec k} = \pi/2$ and $\alpha_{\vec k} = \phi + \pi/2$,
  then this formula reduces to
  $ i e^{ - i \phi \mathbf{\tau_3}}$ we have met before.  Generally, we note
 now ${\rm det} \mathbf{T} (\phi) = -1$ independent of $\phi$, so
 we can take for example $ \mathbf{G}(\phi) = -i \mathbf{T} (\phi)$.
 Using eq (\ref{tr}), we find that $\mathbf{T}(\phi)$, hence
 $\mathbf{G}(\phi)$, is antiperiodic when $\phi \to \phi + \pi$,
 as ${\rm cos} (\frac{\beta_\infty + \beta_0}{2})$ and
  $e ^ { \pm i (\alpha_\infty + \alpha_0)/2}$ both change sign,
 showing that again the state is topologically non-trivial.

 Let us also examine $P  (\vec k) \equiv
 < \mathbf{\Psi_-^C} (\vec k) | \Theta | \mathbf{\Psi_+^C} (\vec k)
 >$ with $C \to C1$ or $C2$.  Again $|P(\vec k)| = 1$ at
 $\vec k = 0$ or $\infty$.  If $\Delta_-(k_{F-}) < 0$ for all
 $\phi$, we can show that again $P(\vec k)$ must vanish
 somewhere in $\vec k$ space.  Explicit calculations show that $P(\vec k)$ is
 real, so it still must vanish on a line in $\vec k$ space.
 For $k_{F+} (\phi) = k_{F-} (\phi) \equiv k_F(\phi)$, this
 happens exactly on the Fermi surface (which in general is
 not a circle).

\section{Conclusion}

To conclude, we have considered the topology of two-dimensional
time-reversal symmetric gapped superconductors by directly analyzing
the quasiparticle wavefunctions, specifically the transformation
function between wavefunctions defined in different $\vec k$ space
regions. This transformation function allows us to classify these
superconductors into two types.  For weak-coupling superconductors,
we verified that the relative sign of the order parameter on the
Fermi surfaces is the crucial parameter which determines the
topology, providing a deeper understanding of earlier results on
vortex bound states \cite{VBS} and edge states \cite{Tanaka09} in
these systems.

\section{Acknowledgement}

 This research was supported by the National Science Council of
 Taiwan under grant number NSC98-2112-M001-019-MY3.

 \appendix

 \section{}

 Here we explain the formulas eq (\ref{T1}), (\ref{T2}) used in
 text.  Suppose that we have a well-defined pseudospin basis
 $| \mathbf{\Psi^{D1}}(\vec k) > $ for $k \ge k_1$, and another
 $| \mathbf{\Psi^{D2}} (\vec k) > $ for $k_2 \ge k \ge 0$, here $k_1 \ge
 k_2$.  Suppose further that we have a solution
 (but not necessarily satisfying eq (\ref{ps1}) and (\ref{ps2}))
  $| \mathbf{\Psi^{C}} (\vec k) > $
 to the Hamiltonian for $k_1 \ge k \ge k_2$.  Then we consider
 the matrix
 \be
 \mathbf{T}(\phi, k_1, k_2)
 \equiv < \mathbf{\Psi^{D1}}(\phi) | \mathbf{\Psi^C}(\phi) >_{k1}
 e^{ -i \Delta \alpha(\phi) \tau_3 / 2 }
  < \mathbf{\Psi^{C}}(\phi) | \mathbf{\Psi^{D2}}(\phi) >_{k2}
  \label{TG}
  \ee
  Here $\Delta \alpha \equiv \alpha(\phi,k_1) - \alpha(\phi,k_2)$,
  where $\alpha(\phi, k)$ was defined in section \ref{sub:so}.
  The matrix $\mathbf{T}$ is in the same spirit as
  eq (\ref{t-def}) except that we now examine the transformation
  between the two pseudospin bases
  $  | \mathbf{\Psi^{D1}} >$ and $| \mathbf{\Psi^{D2}} >$ via an intermediate
  $ | \mathbf{\Psi^C} > $.  The factor
  $ e^{ -i \Delta \alpha(\phi) \tau_3 / 2 }$ is to compensate
  for possible variations in $\alpha_{\vec k}$ between $k=k_1$ and
  $k=k_2$.  In this formula, we have limited ourselves to the
  case where $\beta_{\vec k}$ never equals to $0$ or $\pi$.
  This guarantees that the factor  $ e^{ -i \Delta \alpha(\phi) \tau_3 / 2 }$
  is well-defined despite that $\alpha$ is defined only up to mod $ 2 \pi$:
    For a given $k$, if say $\alpha (\phi, k)$
  increases by $\pi$ when $\phi$ advances by $\pi$, then the same
  thing must happen for a neighboring $k'$ since $\alpha (\phi, k)$
  is smooth (instead of, e.g., $\alpha(\phi, k')$ decreasing by $\pi$).
    Continuing this argument shows that
  $\alpha (\phi + \pi, k_1) - \alpha (\phi, k_1) =
    \alpha (\phi + \pi, k_2) - \alpha (\phi, k_2) $, so
    that $\Delta \alpha (\phi + \pi) = \Delta \alpha (\phi)$
    and $ \Delta \alpha (\phi + 2 \pi) = \Delta \alpha (\phi)$
    without any $ 2 \pi$ ambiguities.
    The precise form of this factor is dependent on the behavior
    of the helicity basis $|\mathbf{\Psi^C}>$ under time reversal
    (see eq (\ref{trC}) below).  A more complicated form is
    necessary for example if we choose to multiply eq
    (\ref{PsiCC1+})-(\ref{PsiCC2-}) by $\vec k$ dependent phase
    factors, which we choose not to do here.

    For simplicity, we have also written eq (\ref{TG}) on two
    circles $k_1$ and $k_2$ in $\vec k$ space.  Since the system
    may not be cylindrically symmetric, one can also choose to
    write this equation on more general paths in $\vec k$ space
    enclosing the orgin,
    but this would complicate the notation.  In the text
    we have only used this formula for $k_1 \to \infty$ and
    $k_2 \to 0$.

    $\mathbf{T}(\phi, k_1, k_2)$ is obviously unitary.
    Let us first write
    \bea
    \mathbf{t^{D1,C} (\phi)} &\equiv& < \mathbf{\Psi^{D1}}(\phi) |
     \mathbf{\Psi^C}(\phi)> _{k1} \nonumber \\
     &=& e^{ i \chi^{D1,C}}(\phi) \mathbf{G^{D1,C}}(\phi)
     \eea
     where ${\rm det} \mathbf{G^{D1,C}} = 1$ etc.  We thus have
     \bea
     \mathbf{T}(\phi) &=& e^{ i (\chi^{D1,C}(\phi)+ \chi^{C,D2}(\phi))}
      \mathbf{G^{D1,C}}(\phi)
     e^{ -i \Delta \alpha(\phi) \tau_3 / 2 }
     \mathbf{G^{C,D2}}(\phi)
       \nonumber \\
       & \equiv &  e^{ i \chi(\phi)} \mathbf{G}(\phi)
     \label{T-long}
     \eea
     Let us now investigate the consequence of time-reversal symmetry.
      Eq (\ref{ps1}) and (\ref{ps2}) require
      \be
       \left( | \Psi^{D1}_1 (\phi+\pi) > , | \Psi^{D1}_2 (\phi+\pi) >
       \right)_k
             =
       \left( |\Theta \Psi^{D1}_1 (\phi) > , |\Theta \Psi^{D1}_2 (\phi) > \right)_k
               \left( \ba{cc}  0 & 1 \\
                              -1 & 0  \ea \right)
      \ee
      For our helicity basis, we have, from eq (\ref{PsiCC1+})-(\ref{PsiCC2-})
      \be
       \left( | \Psi^{C}_+ (\phi+\pi) > , | \Psi^{C}_- (\phi+\pi) >
       \right)_k
             =
       \left( |\Theta \Psi^{C}_+ (\phi) > , |\Theta \Psi^{C}_- (\phi) >
       \right)_k
               \left( \ba{cc}   - e^{ i \alpha(\phi,k)} & 0 \\
                              0 & - e^{ - i \alpha (\phi,k)}  \ea \right)
       \label{trC}
      \ee
      We thus obtain
      \be
      \mathbf{t^{D1,C}}(\phi + \pi) =   \left( \ba{cc}  0 & -1 \\
                              1 & 0  \ea \right)
               \left( \mathbf{t^{D1,C}}(\phi) \right)^*
                   \left( \ba{cc}   - e^{ i \alpha(\phi,k_1)} & 0 \\
                              0 & - e^{ - i \alpha (\phi,k_1)}  \ea \right)
                              \ee
    Similarly,
          \be
      \mathbf{t^{C,D2}}(\phi + \pi) =
                   \left( \ba{cc}   - e^{ - i \alpha(\phi,k_2)} & 0 \\
                              0 & - e^{  i \alpha (\phi,k_2)}  \ea \right)
                    \left( \mathbf{t^{C,D2}}(\phi) \right)^*
                     \left( \ba{cc}  0 & 1 \\ -1 & 0  \ea \right)
                              \ee
   By judicially inserting $ 1 =  \left( \ba{cc}  0 & 1 \\ -1 & 0  \ea \right)
    \left( \ba{cc}  0 & -1 \\ 1 & 0  \ea \right) $ and using
    the commutation properties of Pauli matrices, we find
      \be
     \mathbf{T}(\phi + \pi) =
     e^{ -i (\chi^{D1,C}(\phi) +\chi^{C,D2}(\phi))}
     \mathbf{G^{D1,C}}(\phi)
     e^{ -i \Delta \alpha(\phi) \tau_3 / 2 }  \mathbf{G^{C,D2}}(\phi)
     \ee
    where we have used
     $\Delta \alpha (\phi + \pi) = \Delta \alpha(\phi)$.
     We then obtain
     \be
     e^{ i \chi (\phi + \pi)} \mathbf{G}(\phi + \pi)
      =  e^{ - i \chi (\phi)} \mathbf{G}(\phi)
      \label{T-final}
      \ee
      thus allowing classifications by the behavior of
      $\mathbf{G}(\phi)$ under $\phi \to \phi + \pi$
      as in section \ref{sub:planar}.

    Eq (\ref{TG}) is directly related to the transformation
    used in \cite{Roy06,Kane06} and formula such as eq (\ref{t-def}).
    The analogy to these would be to study
    \be \mathbf{t^{D1,D2}} (\phi) = < \mathbf{\Psi^{D1}}(\phi) |
     \mathbf{\Psi^{D2}}(\phi)>_K
     \label{tDD}
     \ee
      at some $K$ where
     both $ |\mathbf{\Psi^{D1}}(\vec k)>$ and
     $|\mathbf{\Psi^{D2}}(\vec k)>$ are well-defined.
     If indeed $  |\mathbf{\Psi^{D2}}(\vec k)>$ in eq (\ref{TG})
     is defined for $k$ up to $k_1$, then we see that, using that
      $  |\mathbf{\Psi^{D2}}(\vec k)>$ forms a complete set for
      the positive energy levels,
     \be
     \mathbf{T}(\phi) =  < \mathbf{\Psi^{D1}}(\phi) |
     \mathbf{\Psi^{D2}}(\phi)>_{k_1} \mathbf{M} (\phi, k_1, k_2)
     \ee
     where
     \be
      \mathbf{M} (\phi, k_1, k_2) =
      < \mathbf{\Psi^{D2}}(\phi) | \mathbf{\Psi^{C}}(\phi)>_{k_1}
            e^{ -i \Delta \alpha(\phi) \tau_3 / 2 }
      < \mathbf{\Psi^{C}}(\phi) | \mathbf{\Psi^{D2}}(\phi)>_{k_2}
      \label{M}
      \ee
      so amounts evaluating (\ref{tDD}) at $k_1$ and
      post-multiplying by the matrix $\mathbf{M}$.  By the
      same analysis as from eq (\ref{TG}) to (\ref{T-final}),
      we see that the SU(2) part of $\mathbf{M}$ must be
      either periodic or antiperiodic when $\phi \to \phi + \pi$.
      However, since the wavefunctions entering $\mathbf{M}$ are
      smooth and  $\mathbf{M}(\phi,k_2,k_2)$ is necessarily
      simply $1$, the SU(2) part of $\mathbf{M}(\phi,k_1,k_2)$
      must be periodic when $\phi \to \phi + \pi$.
      Hence the classification is not affected
      by the post-multiplication by this matrix.

      Similarly, one can show that the classification by the SU(2)
      part of (\ref{TG}) is independent of the choice of
      the pseudospin basis $|\mathbf{\Psi^{D1}}>$ or
        $|\mathbf{\Psi^{D2}}>$, so long as they are smoothly
        defined in their respected regions in $\vec k$ space.
        A different choice would just amount to pre- or
        post-multiplying eq (\ref{TG}) by a matrix with the
        SU(2) part of which being periodic in $\phi \to \phi + \pi$.

      We have assumed that $\beta_{\vec k}$ is never $0$ or $\pi$
      in eq (\ref{TG}).  When this does not apply,
      the factor $ e^{ -i \Delta \alpha(\phi) \tau_3 / 2 }$
      there has to be rewritten.  The
      necessary form can be obtained by considering more carefully
      how eq (\ref{trC}) changes when $\vec k$ is varied,
       but we shall not go into these complications
      here.

      When there are sign changes of $\Delta_-(\vec k)$ for both
      outside and inside $k_{F-}$, one must use
      $  | \mathbf{\Psi^{C1}}(\phi) > $ and $  | \mathbf{\Psi^{C2}}(\phi)
      >$ for these two regions separately.  Similar considerations
      above show that a transformation matrix which can be used is
   \bea
       \mathbf{T}(\phi, k_1, k_2)
 &\equiv& < \mathbf{\Psi^{D1}}(\phi) | \mathbf{\Psi^{C1}}(\phi) >_{k1}
 e^{ -i \Delta \alpha_1(\phi) \tau_3 / 2 }
 < \mathbf{\Psi^{C1}}(\phi)| \mathbf{\Psi^{C2}}(\phi)>_{k_{F-}(\phi)}
   \nonumber \\
   & &  \qquad \times  e^{ -i \Delta \alpha_2(\phi) \tau_3 / 2 }
  < \mathbf{\Psi^{C2}}(\phi) | \mathbf{\Psi^{D2}}(\phi) >_{k2}
  \label{TG2}
  \eea
    where $\Delta_1 (\phi) \equiv \alpha(\phi,k_1) -
    \alpha(\phi,k_{F-}(\phi))$ and
    $\Delta_2 (\phi) \equiv  \alpha(\phi,k_{F-}(\phi)) -
    \alpha(\phi,k_2)$.  If $\alpha$ is independent of
    $k$, this formula reduces to eq (\ref{TR2}) at
    the end of section \ref{sub:Rashba}.


\begin{thebibliography}{plain}
\bibitem{note3d}  Tological insulators can also exist in
dimensions other than two\cite{Kane07,Ludwig08}.  In this paper
however, we shall confine ourselves exclusively to two-dimensions.
\bibitem{wc} By this we mean that the chemical potential is
positive.  For this condition, see for example \cite{Read00} and
discussions in text.
\bibitem{unocc}  In the topological insulators literature
\cite{Kane05,Kane06,Kane07}, the topology is studied for the
occupied states.  Here, we choose instead to study the unoccupied
states.  These are equivalent since the occupied and unoccupied
states always come in positive and negative energy pairs in
superconducting systems.
\bibitem{exp}  We have removed the plane wave factors $e^{ i \vec k \cdot \vec
r}$ of these wavefunctions before discussing the topology, as also
done in e.g. \cite{Kane05,Kane06,Kane07,Kohmoto85}.  We however
would not introduce another symbol for simplicity in notations.
\bibitem{T} The wavefunction of the time reversed state can be
obtained by complex conjugation and then pre-multiply by the matrix
\be \left( \ba{cccc} 0 & -1 & 0 & 0 \\
                     1 & 0  &  0 & 0 \\
                     0 & 0  & 0 & -1 \\
                     0 & 0 & 1 & 0 \ea \right) \ee
Note $\phi_{-\vec k} = \phi_{\vec k} + \pi \ {\rm mod} \ 2 \pi$.
\bibitem{infty}  When discussing the topology, we regard $\vec k \to \infty$ as a single point irrespective
of its direction, as done in \cite{Read00} etc.   We shall therfore
require well-defined pseudospin wavefunctions to have $\phi$
independent values as $\vec k \to \infty$. This condition would not
be imposed upon on wavefunctions such as $|\mathbf{\Psi^C}>$ in
subsections \ref{sub:Rashba} and \ref{sub:so} that are used as
intermediate wavefunctions.
\bibitem{transpose}  Strictly speaking we need a transpose here, but
we shall suppresses this to simplify notations.
\bibitem{ambi}  $\mathbf{G}(\phi)$ is still ambiguous up to
a $\pm$ sign.  However, this does not cause any concern since we
would be studying a smooth map $ \phi \to \mathbf{G}(\phi)$.
\bibitem{even} $|P(k)|$ must be unity at $\vec k = 0$ and $\vec k= \infty$.
In the language of \cite{Kane06},  these are time-reversal invariant
points and must long to the "even subspace" of $P(k)$.
\end{thebibliography}
\end{document}